\documentclass[sigconf,edbt]{acmart-edbt2024}
\pdfoutput=1
\def\BibTeX{{\rm B\kern-.05em{\sc i\kern-.025em b}\kern-.08em
    T\kern-.1667em\lower.7ex\hbox{E}\kern-.125emX}}

\usepackage{booktabs} 

\setcopyright{rightsretained}

\acmDOI{}

\acmISBN{978-3-89318-091-2}

\acmConference[EDBT 2024]{27th International Conference on Extending Database Technology (EDBT)}{25th March-28th March, 2024}{Paestum, Italy}
\acmYear{2024}

\usepackage{moresize}
\usepackage{multirow}
\usepackage{multicol}
\usepackage{pifont}
\usepackage{array}
\usepackage{url}

\usepackage{comment}
\usepackage{pgfplots} 
\usetikzlibrary{patterns}
\usepackage[caption=false]{subfig}
 \pgfplotsset{compat=1.18}
 \usepackage{mathrsfs}
 \usepackage{balance}
 \usepackage{blindtext}
\usepackage{amsthm}

\theoremstyle{definition}

\begin{document}
\title{Data-CASE: Grounding Data Regulations for Compliant Data Processing Systems}
  
\author{Vishal Chakraborty}
\email{vi.c@uci.edu}
\affiliation{%
  \institution{University of California (UC) Irvine}
}
\author{Stacy Ann-Elvy}
\email{selvy@ucd.edu}
\affiliation{%
  \institution{UC Davis School of Law}
}

\author{Sharad Mehrotra}
\email{sharad@ics.uci.edu}
\affiliation{%
  \institution{UC Irvine}
}
\author{Faisal Nawab}
\email{nawabf@uci.edu}
\affiliation{%
  \institution{UC Irvine}
}

\author{Mohammad Sadoghi}
\email{msadoghi@ucdavis.edu}
\affiliation{%
  \institution{UC Davis}
}

\author{Shantanu Sharma}
\email{shantanu.sharma@njit.edu}
\affiliation{%
  \institution{New Jersey Institute of Technology}
}
\author{Nalini Venkatasubramanian}
\email{nalini@uci.edu}
\affiliation{%
  \institution{UC Irvine}
}

\author{Farhan Saeed}
\email{fsaeed1@uci.edu}
\affiliation{%
  \institution{UC Irvine}
}

\renewcommand{\shortauthors}{}

\begin{abstract}
Data regulations, such as GDPR, are increasingly being adopted globally to protect against unsafe data management practices. Such regulations are, 
 often ambiguous (with multiple valid interpretations) when it comes to defining the expected dynamic behavior of data processing systems. This paper argues that it is possible to represent regulations such as GDPR formally as invariants using a (small set of) data processing concepts that capture system behavior. When such concepts are \emph{grounded}, i.e., they are provided with a single unambiguous interpretation,  systems  can achieve compliance by demonstrating that the system-actions they implement maintain the invariants (representing the regulations). To illustrate our vision, we propose Data-CASE, a simple yet powerful model that (a) captures key data processing concepts
 (b) a set of invariants that describe regulations in terms of these concepts. 
 We further illustrate the concept of grounding using "deletion" as an example and highlight several ways in which end-users, companies, and software designers/engineers can use Data-CASE.
\end{abstract}

%
%


\maketitle
\section{Introduction}
\label{sec: intro}
The rise in organizations collecting and mishandling personal data has led to the emergence of data regulations across the world. 
Examples include the California Consumer Protection Act (CCPA) \cite{ccpa}, the Virginia Data Protection Act (VDPA) \cite{vdpa}, and Canada's Personal Information Protection and Electronic Documents Act (PIPEDA) \cite{PIPEDA}. 
Many countries are in the process of enacting their own laws. Of these, the most developed, scrutinized, and used is ``The Regulation (EU) 2016/679'' also known as the  {\em General Data Protection Regulation (GDPR)} \cite{GDPROffText}. The Data Governance Act adopted by the EU in 2022 complements the GDPR \cite{DGA}. 

Building systems that allow compliant data governance has been identified as a key challenge in the recent Seattle Report on Database Research\cite{SeattleReport22}. While GDPR has resulted in tangible improvements in how organizations handle data \cite{WhatsAppChange, FBInstaChange}, it has nonetheless led to negative economic impacts, one of the causes for which is the uncertainty companies face in ensuring compliance \cite{CDINeg, LegalUncert} and the risk of penalty if found non-compliant \cite{NYTadDrop, EuropeadDrop, shastri2019seven,ruohonen2021gdpr}. 
A key reason for such uncertainty is the ambiguity in the legal language used in data regulations  when it comes to how systems should process data, i.e., which actions
should the system take and when to comply with data regulations. Many of the concepts
listed are open to several valid interpretations.  

Consider a company MetaSpace that stores personal data of individuals, including their location for smart space applications, using PostgreSQL (PSQL). A user wishes to exercise their right (GDPR \textsc{Article} 17) to have their data deleted in a reasonable time. Persistently deleting data, e.g., using VACCUM-DELETE in PSQL can be extremely expensive. On the other hand, adopting logical deletes as in Cassandra, a NoSQL distributed database \cite{Cassandra}, --- inserts a \emph{tombstone} when data is deleted--- can be efficient. Prior work \cite{Lethe} has shown that using delete markers like tombstones in LSM trees may lead to data being, illegally, physically retained for a long duration. The impact of the ambiguity is further highlighted when we consider distributed systems that may replicate /cache data across different nodes using various complex protocols \cite{ConFMO}. If erasure means removing the data not just from the primary location, but removing it completely  (from all locations in disk and memory),
 a technique will have to be built to track the copies and delete all of them. Thus, due to the lack of system specifications, ambiguities arise which can expose the company to legal action.

    \begin{figure*}[t]
    \centering
        \hspace{-1cm}
         \includegraphics[page=5, trim = 3mm 90mm 5mm 52mm, clip, width=18cm]{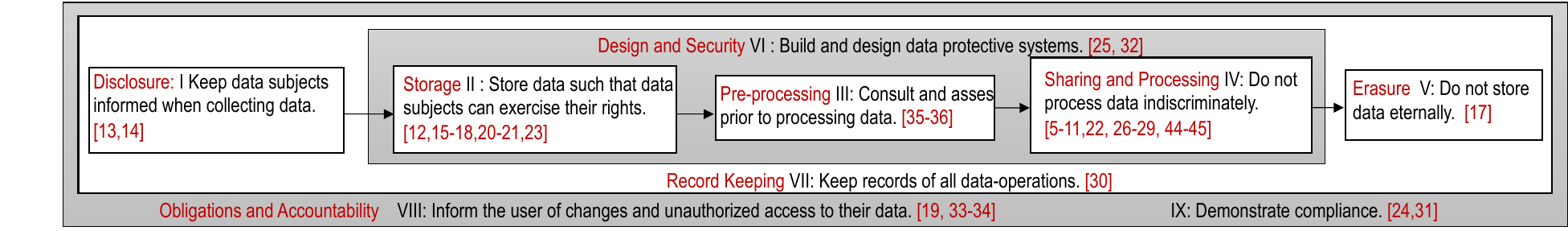}
        \caption{The GDPR requirements for data governance stated as informal invariants and the grouped articles.}
        \label{fig:cmd_art}
    \end{figure*}

\color{black}
\begin{figure}[t]
    \centering
     \includegraphics[page=10, trim = 115mm 83mm 105mm 52mm, clip, width=6cm]{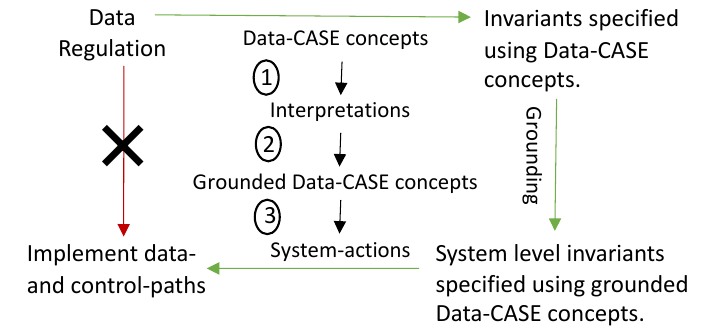}
        \caption{Schematic representation of Data-CASE}
    \label{fig:Data-CASE_OverView}
\end{figure}

Ambiguity in interpreting GDPR has been a cause of concern for the research community as evidenced by the work of the Article 29 Data Protection Working Party (AWP29) \cite{WP29} and EDPB \cite{EDPBClar}, 
which issues clarifications and recommendations on how organizations may attain compliance. These reports, however, clarify only limited aspects of the  GDPR regulation. Furthermore,  prior work~\cite{doi:10.1073/pnas.1914598117} has 
shown that, at times, AWP29's clarifications and recommendations have been 
unsound \cite{PDAWP29,doi:10.1073/pnas.1914598117} and do not meet the desired compliance. 

\color{black}{
To bridge the gap between ambiguous legal specifications and grounded (system-level) technical specifications that can serve
as a blueprint for compliance in systems, 
we need a model with a set of concepts that can be used to translate  
the data governing requirements in regulations into a set of well-defined specifications of dynamic system behavior.
Such a model must: (a) consist of a set of (data processing) concepts to fully describe data regulations; (b) allow different valid interpretations of the regulations; (c) allow for unambiguous interpretations of the concepts to be mapped to system-actions 
using which they can be implemented in a system.

 Using such a model, organizations working along with regulatory bodies 
 can agree upon possible desirable interpretations of concepts/properties and attain demonstrable compliance.

Towards the goal of developing such a model, we propose Data-CASE which stands for  \textit{Data Collection, Access, Sharing, and Erasure} model. Data-CASE consists of a small set of key system  concepts (e.g., erasure, encryption, policies), referred to as \emph{Data-CASE concepts}. 
Concepts in Data-CASE are chosen such that the specifications in regulations that relate to the requirements of systems can be expressed formally using these concepts as invariants in a logic framework. 
Each Data-CASE concept  may have several possible interpretations, e.g., deletion/erasure may have
different interpretations as 
discussed earlier.
Data-CASE allows for such interpretations to be formally defined 
 (\textcircled{1}, Figure \ref{fig:Data-CASE_OverView}). Furthermore, the system developers/deployers can choose the specific 
 interpretation of the concepts they wish to support in their system (\textcircled{2}, Fig. \ref{fig:Data-CASE_OverView}). 
 Such an interpretation is then finally mapped to system-level actions --- \emph{system-actions} --- (\textcircled{3}, Fig. \ref{fig:Data-CASE_OverView}) to achieve the specified
 concept interpretation.  We refer to this process of choosing a specific interpretation for a given concept and formally specifying the chosen interpretation as the process of \emph{grounding} and refer to the formally defined concepts as the \emph{grounded} interpretations. Note that the system-action to which the grounded concept is mapped is system dependent. In cases where a system-action is not supported to exactly implement the chosen interpretation, the system might need to be retrofitted or changed to support the necessary system-actions. \textcolor{black}{Examples of system-actions include DELETE and VACUUM in PSQL; deleteOne and remove in MongoDB. System-actions may also include user defined functions. Note that Data-CASE is neither a system nor system-dependent. It is a formal framework that can be used for reasoning about compliance in any system by creating mappings between grounded concepts and system-actions.}

\textcolor{black}{\textsc{Organization.} In Section (\S) 2, we propose a basic model for Data-CASE focusing on only key concepts we think such a model must contain. 
In \S3, we show how concepts defined in \S2 can be grounded to remove 
ambiguity by demonstrating, as an example, how valid interpretations of data erasure can be formally stated using Data-CASE.
In \S4 we show some use cases of Data-CASE.
 \S5 discusses related work. \S6 highlights the challenges towards the goal of a fully specified model that can abstract compliance for data processing system development and deployment. 
}

\section{Data-CASE Model} 
\label{sec: Model}
Data regulations specify the kind of data that falls under their domain and legislate how such data is handled as it flows through a data processing system and the responsibilities of entities processing such data. So, Data-CASE groups the requirements of a given data regulation under the following eight categories (the first five corresponding to the data life cycle and the remaining three to system properties) ---(1) Disclosure, (2) Storage, (3) Pre-processing, (4) Sharing and Processing, and (6) Erasure, (7) record keeping, and (8) obligations and accountability. We illustrate this schematically in Figure \ref{fig:cmd_art} where we also group the articles of GDPR \textcolor{black}{(only those that legislate data processing and impact system design \cite{shastri2019understanding})} under these categories.

To capture the data governing principles in a data regulation, the first step is to logically differentiate personal data from other kinds of data. In addition, it is essential to capture the provenance between various kinds of data in a system. Since data regulations grant rights to owners of personal data,
there is a need to support user policies, track their evolution over time, and validate them. Finally, when data is accessed/used, it is done so for specific purposes by specific entities which are restricted by policies set by the owner of such data or the regulation.
Below, we develop a simple set of concepts that capture the above.


\subsection{Data Processing Concepts in Data-CASE}
\label{subsec: Concepts}
 As data flows through the data-life cycle, it is collected from the data-subject by the controller who might share it with processors. Auditors verify and certify compliance. In Data-CASE, these roles are referred to as entities. We denote them with $e$. As a running example, consider Netflix collects the credit card information of its subscribers and stores it on the AWS cloud.
 
 The concept of \textbf{data unit} refers to the finest granularity 
at which we refer to data in Data-CASE. 
The granularity depends on the system, application, as well as data regulation. 
E.g., in a website collecting info about click-stream, a specific user's data might
be a data unit. In a sensor such as a camera, the unit might correspond to all data
(irrespective of who is in it) of a camera within a certain interval. For a service provider like Netflix, credit card information can be considered a data unit.

We denote a data unit as a tuple  $X=(S, O, V, P)$ where $S$ is the data-subject---the entity whom the data identifies; $O$ is the origin---where the data was collected from; $V$ is a set $\{(v_1,t_1), (v_2, t_2), \hdots\}$ of values where $v_i$ is the value at time $t_i$, and $P$ is the set of associated policies. A collection of data units is denoted as $\vec{X}.$ 

Data-CASE classifies data units into three categories --- (1) \textbf{base data}, which is directly or indirectly collected, (2) \textbf{derived data}, which is obtained from base data, and (3) \textbf{metadata}, which includes data-subject, policies, etc. 
A \textit{derived} data unit has the same four aspects as base data except that data-subject (and origin) are possibly varying sets of the data-subjects (and origins) of the base data from which it was derived. The aspects of the derived data unit are aggregated over the aspects of the base data.

A task or service, for which collected data is used, identifies its \textbf{purpose} of data processing. Collected base data can have more than one purpose. E.g., Netflix collects credit card information for billing; saves view-history for targeted advertisements, etc. 
\color{black}
\\
The flow of data units through various stages of processing in a system is controlled using policies. A \textbf{policy} on a data unit $X$ is a tuple $\langle p, e, t_b, t_f \rangle$ is a constraint specifying that an entity $e$ can access the data unit for purpose $p$ from time $t_b$ to $t_f.$ 
\\
\color{black}
\noindent\textsc{Example:} The policy $\pi_1 = \langle \texttt{billing}, Netflix,$ $ 010123,01012024 \rangle$ for a data unit $X=credit\_card$  of user $1234$ states that Netflix has access to $X$ for the purpose of billing from 01/01/23 through 01/01/24. The policy $\pi_2 = \langle \texttt{retention}, AWS,$$ 010123,010124\rangle$ on $X$ states that AWS can retain this data from 01/01/23 through 01/01/24.
\color{black}
For a data unit $X$, we write $V(t)$ to denote its value at time $t$ and $P(t) := \{(p,e, t_b, t_f) \in P \mid t_b \leq t \leq t_f \}$ to denote the set of policies on $X$ at time $t$. The \textbf{state} of a data unit $X$ at a given time are the values of its aspect at that time and are denoted as $X(t)= (S(t), O(t), V(t), P(t))$. The state of a database is the collection of the states of all data units in the database.
\\
\color{black}
\textsc{Example:} The state of $X$ in the previous example at time $t =$ 00:10 on 02/26/23 is $X(t) = (1234, 0, credit\_card\_info, \{ \pi_1, \pi_2, \ldots \}).$
\color{black}

We refer to any operation that changes the state of data units as an \textbf{action}. 
Actions include the creation and deletion of data units, changes to the value of a data unit, and reads and writes on any aspect of a data unit. An action can influence one or more data units. For an action $\tau$ on data unit $X$, we denote the changed state of $X$ with $\tau(X).$
Actions on data units in a database $D$ give rise to a series of states $\mathcal{D}_1, \mathcal{D}_2, \hdots.$ 
Actions can produce derived data. A derived data unit $Y = (S_Y, O_Y, V_Y, P_Y)$ produced by action $\tau$ from a collection of base data units $\vec{X}$ has $S_Y$ and $O_Y$ as the union of all the data-subjects and origins of data units $\vec{X}$, respectively. The set of policies $P_Y$ is generally a restriction of the policies of the data units in $\vec{X}.$

Data regulations often require monitoring how data is processed or changes over time. Each action on a data unit is denoted as an \textbf{action-history tuple}. A collection of action-history tuples is called an \textbf{action-history.} 
For a data unit $X$, and a database $D$,
\\
$\bullet$ an action-history tuple is given by $(X, p, e, \tau(X), t)$ denoting that entity $e$ performed action $\tau$ on $X$ for purpose $p$ at time $t.$ 
\\
$\bullet$ action-history of $X$ denoted, $\mathcal{H}(X)$, is the set of all actions on $X$, i.e., $\mathcal{H}(X) = \{ (X, p, e, \tau_i(X), t)_i\}_{i=1}^n.$

\noindent\textsc{Example:} The action-history tuple
$(1234, \texttt{comp}, Netflix, \texttt{CtrC1234},$ $ 010223)$ records that
on 01/02/23, Netflix made a contract to collect data of user $1234$. Such a contract gets the consent of the user to set policies $\pi_1$ and $\pi_2$ in previous examples. Similarly, the tuple $(X, \texttt{billing}, Netflix, \texttt{read}(credit\_card), 0010-022623)$ records that Netflix accessed the credit card information of $1234$ for billing at 00:10 on 02/26/23.

\color{black}
Data regulations specify what constitutes lawful data processing. Data-CASE abstracts lawful data processing as \textbf{policy-consistent} data processing.
For a data unit $X = (S,O, V, P)$, action $\tau$ on $X$ for purpose $p$ at time $t$, we say that the action-history tuple $(X, p, e, \tau(X), t)$ on data unit $X$ is \emph{policy-consistent} if there exists a policy $\langle p, e, t_b, t_f \rangle$ in $P(t)$ in the state $(S, O, V(t), P(t))$ of data unit $X$ or the action in the tuple is required by a data regulation. We say that actions on $X$ are policy-consistent if every action-history tuple in $ \mathcal{H}(X)$ is policy-consistent. 


\subsection{Formal Invariants For Compliance}
\label{subsec: invariants}
Having described the set of concepts above (data unit, policy-consistent action, etc.), we can now specify data regulations formally in the form of invariances. We provide two examples. 

GDPR \textsc{Article} (denoted $\mathcal{G}$) 6 defines when processing personal data is lawful. Data-CASE abstracts this notion using the concept of \emph{policy-consistent data processing}. It can be stated as: $\text{ For all data units }X, \text{ and }$ $\text{for all actions }$ $ \tau \text{ on }X, \text{ it holds that } \tau$ is policy-consistent. The legally permissible grounds and data-subject's consent for processing data can be encoded as specific purposes through policies in Data-CASE.

Consider $\mathcal{G}17$. It requires that personal data be not retained longer than necessary for the purpose they were collected and they be deleted without undue delay. Formally, this can be specified as follows. For all data units $X = (S, O, V, P)$, there exists a policy $\pi \in P$ such that $\pi = \langle \texttt{compliance-erase}, e, t_b, t_f \rangle$ and the last access tuple on $X$ is $(q, \texttt{compliance-erase}, e, \texttt{erase}(X), t)$ s.t. $t \leq t_f.$
The above statement states that every data unit $X$ has a policy associated with it, which states that the data unit has to be erased due to compliance requirements at a specified time. Moreover, the last action on the data unit $X$ is $\texttt{erase(X)}$ at a time earlier than the time within which the policy requires the data unit to be deleted.


Observe that the above invariants capture $\mathcal{G}$6 and $\mathcal{G}$17 formally. However, the Data-CASE concepts \emph{erasure} and \emph{policies} are still open to multiple valid interpretations. In the next section, we discuss the process of grounding such concepts so that they can be mapped to specific implementations in systems. 


\section{Grounding Concepts}
\label{sec: Grounding}
The process of grounding consists of mapping a concept to a unique interpretation and formalizing it in Data-CASE. We illustrate the process through \emph{erasure}. Recent work \cite{Lethe,sarkar:hal-01824058,sarkar2022query} has shown that many complexities arise when interpreting and implementing erasure and can have a significant impact on system performance. Besides, erasure is a requirement of most regulations, and has received considerable attention recently \cite{TikTokUSCong, TikTokUSCongDel}.

\subsection{Data erasure}
\label{subsec: Erasure Grounding}
In a system, erasure can be interpreted in various ways. 
We consider four interpretations - inaccessibility, deletion, strong deletion, and permanent deletion.
\\
    $\bullet$ We say that data is \emph{reversibly inaccessible} in a system when it cannot be read by any data-subjects in the system but remains accessible to the controller or processor. 
    Often, in such cases, it can be accessed by the data-subject after a specific action. 
\\
    $\bullet$ We say that data has been \emph{deleted} when the data and all its copies have been physically erased. 
\\    
    $\bullet$ We say that data has been \emph{strongly deleted} when it has been deleted and all dependent data, where the data-subject is identifiable, has been deleted. 
\\  
    $\bullet$ We say that data has been \emph{permanently deleted} when it has been strongly deleted, and some advanced physical drive sanitation technique has been used. 

Observe that these interpretations can be ordered based on their restrictiveness. For example, strongly delete implies delete. This gives rise to the notion of strictness of interpretation of compliance.
Figure \ref{fig:datatimel} depicts the temporal relationship between the different interpretations.  While some notions of erasure are arguably better than others when privacy and security of personal data are considered, these deletion methods have different overheads and may or may not be considered practical or feasible.
To ground these interpretations, we identify three \textcolor{black}{properties}.
\\
$\bullet$ \textit{Erasure-inconsistent read:} (Illegal Reads-IR) We say that there is an erasure-inconsistent read on data unit $X$ if there exists the tuple $(X, q, p, e, \texttt{read}(X), t_j) \in \mathcal{H}(X)$ and in $X(t_j)$ we have $P(t_j) = \emptyset$, i.e., $X$ was read although there were no policies authorizing it.
\\
$\bullet$ \textit{Erasure-inconsistent inference:}(Illegal Inference-II) We say that there is an erasure-inconsistent inference on data unit $X$ if there exists the tuple $(X, q, p, e, \texttt{erase}(X), t_j) \in \mathcal{H}(X)$ and $X = f(Y)$ where $Y$ is other data units and $f$ is some dependency that can be used to reconstruct $X$ from $Y$. i.e., although $X$ has been erased, it can still be inferred using dependent data, provenance data, or from other data units in the database.
\\
$\bullet$ \textit{Transformation invertibility}:(Invertibility-Inv.) A data unit $X$ is usually transformed to some $f(X)$ (e.g., an encryption function, a function that rewrites $X$ with 0 bits, etc.) to prevent illegal reads and illegal inferences. The transformation (function) may be invertible, i.e., recoverable, or non-invertible. 

We can formally ground the four different deletion notions. Table \ref{tab:erasure_notions} characterizes the different notions of erasure in terms of the actions introduced earlier. Observe that although \emph{strong delete} and \emph{permanent delete} have the same properties, the latter entails the additional step of advanced data sanitization like \cite{DoDsanitise}.
\begin{table}[t]
    \centering
    \small
     \caption{Interpretations of erasure and their characteristics. 
     \footnotesize{$\checkmark$ indicates feasibility and $\times$ indicates not. }}
    \begin{tabular}{lcccl}
    \hline
         \textbf{Erasure} & \textbf{IR} & \textbf{II} & \textbf{Inv} & \textbf{PSQL System-Action(s)}  \\
         \hline
         reversibly accessible & $\times$ & $\checkmark$ & $\checkmark$ & Add new attribute \\
         delete & $\times$ & $\checkmark$ & $\times$ & DELETE+VACUUM \\
         strong delete & $\times$ & $\times$ & $\times$ & DELETE+VACUUM FULL\\
         permanently delete & $\times$ & $\times$ & $\times$ & Not supported\\
         \hline
    \end{tabular}
    \vspace{-2pt}
    \label{tab:erasure_notions}
\end{table}
\begin{figure}[t]
    \centering
     \includegraphics[page=2, trim = 25mm 72mm 0mm 88mm, clip, width=9cm]{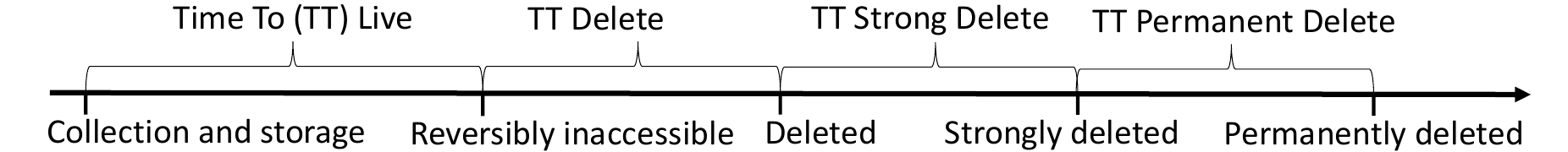}
    \caption{Data Erasure Timeline}
    \label{fig:datatimel}
\end{figure}
\subsection{Other Concepts}
\label{subsec: Other concepts}
\color{black}
Other Data-CASE concepts such as purpose, histories, and policies also need to be grounded when reasoning with systems being compliant with a given data regulation. Grounding concepts require a careful analysis of actions different systems use for these concepts, as well as, interactions between the actions. For example, to ground histories, one has to consider various logs a system maintains, their granularity, and uses --- logs may be temporary or kept for a long duration to not only recover data but also to support the rights of data-subjects. Furthermore, logs directly impact requirements like demonstrating compliance, system recovery, and data erasure. Similarly, purposes need to be grounded to specific actions. A purpose typically calls for a set of authorized actions. E.g. the purpose of billing only allows the credit card information to be read and processed with the bank and not share it with a third party. This, in turn, directly impacts policy-consistent data processing. Finally, each grounded concept needs to be mapped to system-actions.

Thus, grounding all concepts which enable study interactions between them in a given system is a complex task and is fundamental to defining what compliance means for that system.
\color{black}
\begin{figure*}[t]
    \subfloat[Interpretations of Data Erasure in PSQL on WCus]{
\begin{tikzpicture}
\begin{axis} [name = plot1, height=3.86cm,
        width=6cm,
        ymin=0,
        ymax=16000,
        xmin = 0,
        xmax = 13,
        xticklabel shift= 0in,
        xtick={2,5,8,11},
        xtick pos=bottom,
        xticklabels = {10K, 30K, 50K, 70K},
        x tick label style = {font = \small, rotate=0},
        y tick label style = {font = \small, rotate=0},
        ylabel=Completion Time (secs),
        xlabel= {Number of Transactions (txns.)},
        x label style = {font=\small},
        y label style = {font=\small},
        legend style={
          at={(0,1)},
          anchor=north west, font =\small, legend columns = 1, draw=none, fill = none
        },
        ybar=0pt, 
        bar width=6pt,
        legend pos= north west,
        legend style={legend cell align=left, font=\tiny, legend columns=1},
        legend image code/.code={\draw [#1] (0cm, -0.1cm) rectangle (0.2cm, 0.125cm);},
        ytick = {1000, 3000, 6000, 9000, 12000, 15000},
        ymajorgrids, yminorgrids, reverse legend
]
\addplot [fill=teal]  coordinates {(2,376) (5,1003) (8,1472) (11,1966)};
\addlegendentry {DELETE + VACUUM}
\addplot [pattern=horizontal lines, pattern color=red] coordinates {(2,412) (5,1044) (8,1617) (11,2042)};
\addlegendentry {DELETE}
\addplot [pattern=north east lines, pattern color=brown] coordinates {(2,459) (5,1321) (8,2261) (11,3247)};
\addlegendentry {Tombstones (Indexing)}
\addplot [pattern=north west lines, pattern color=black]coordinates {(2,2391) (5,7106) (8,11248) (11,14826)};
\addlegendentry {DELETE and VACUUM FULL}
\end{axis}
\end{tikzpicture}
}
\subfloat[Completion time for workloads]{
\begin{tikzpicture}
\begin{axis} [height=3.86cm,
        width=6cm,
        ymin=0,
        xmin = 0,
        xmax = 8,
        xticklabel shift= 0.0in,
        xtick={1,3,5,7},
        xtick pos=bottom,
        x tick label style = {font = \small, rotate=0},
        xticklabels = {WPros, WCon, WCus, YCSB-C},
        x tick label style = {font = \small, rotate=0},
        ylabel={Completion Time (mins)},
        ytick = {5, 10, 15, 20, 25, 30},
        x label style = {font=\small},
        y label style = {font=\small},
        xlabel= {Workloads (100k records, 10k txns.)},
        legend style={
          at={(10,1)},
          anchor=north east,
        },
        ybar=0pt, 
        bar width=6pt,
        legend pos= north east,
        legend style={draw = none, fill = none, font=\tiny, legend columns=3, at={(1,1)}},
        legend image code/.code={\draw [#1] (0cm, -0.1cm) rectangle (0.2cm, 0.25cm);},
        ymajorgrids,
]
\addplot [fill=teal] coordinates {(1,6.03) (3,6.42) (5,8.8) (7,5.25)};
\addlegendentry {P\_Base}
\addplot  [pattern=north east lines, pattern color=red]coordinates {(1,6.6) (3,9.67) (5,10.85) (7,6.07)};
\addlegendentry {P\_GBench}
\addplot [pattern=north west lines, pattern color=brown] coordinates {(1, 26.5) (3, 16.11) (5, 16.52) (7,7.12)};
\addlegendentry{P\_SYS}
\end{axis}
\end{tikzpicture}
}
\subfloat[WCus (lines) \& YCSB-C(bars) workloads]{
\begin{tikzpicture}
\begin{axis} [height=3.86cm,
        width=6cm,
        ymin=0,
        ymax=100,
        ytick = {0, 20, 40, 60, 80, 100},
        xmin = 0,
        xmax = 10,
        ylabel={Completion Time (mins)},
        ytick = {5, 10, 15, 20, 25, 30},
        y label style = {font=\small},
        legend style={
          at={(0,0.75)},
        },
        ybar=0pt, 
        bar width=6pt,
        legend pos= north west,
        legend style={draw=none, fill=none, font=\tiny,  at={(0,0.75)},anchor=north west, legend columns=2},
        legend image code/.code={\draw [#1] (0cm, -0.1cm) rectangle (0.2cm, 0.25cm);},
        ymajorgrids,
        y label style = {font=\small},
	xtick pos=bottom,
        xtick={},
	height=3.96cm,
        width=6cm,
        axis x line=none,
        xtick style={draw=none},
        axis y line=none,
        ytick style={draw=none},
]
\addplot[fill=teal] coordinates {
	(1, 5.25)
	(3, 10.3)
	(5, 13.47)
	(7, 16.13)
        (9, 19.72)

};
\addplot[pattern=north east lines, pattern color=red] coordinates {
	(1, 6.07)
	(3, 10.98)
	(5, 15.18)
 	(7, 18.75)
	(9, 22.48)
};
\addplot[pattern=north west lines, pattern color=brown] coordinates {
	(1, 7.12)
	(3, 11.5)
	(5, 15.77)
	(7, 18.63)
	(9, 21.55)
};
\legend{P\_Base, P\_GBench, P\_SYS}
\end{axis}
\begin{axis}[
	xlabel=Number of Records (10k txns.),
	ylabel=Completion Time (mins),
	x label style = {font=\small},
        y label style = {font=\small},
	xtick pos=bottom,
	height=3.96cm,
        width=6cm,
        legend style={draw=none, fill=none, font=\tiny,  at={(0,1)},anchor=north west, legend columns=2},
    xticklabels={100k, 200k, 300k, 400k, 500k},
    xtick = {1,2,3,4,5},
    ymajorgrids,
    ]
\addplot[color=teal,mark=x] coordinates {
	(1, 8.8)
	(2, 18.33)
	(3, 27.97)
	(4, 38.63)
	(5, 48.77)
};
\addplot[color=red,mark=square] coordinates {
	(1, 10.85)
	(2, 19.78)
	(3, 31.75)
	(4, 44.32)
	(5, 55.57)
};
\addplot[color=brown,mark=diamond] coordinates {
	(1, 16.52)
	(2, 33.58)
	(3, 54.53)
	(4, 72.73)
	(5, 95.5)
};
\legend{P\_Base, P\_GBench, P\_SYS}
\end{axis}
\end{tikzpicture}
}
    \vspace*{-15pt}
    \caption{(a) Impact of different interpretations of data erasure on PSQL. (b) Completion time. (c) Scalability. }
    \label{fig:Eval_WkLds}
\end{figure*}
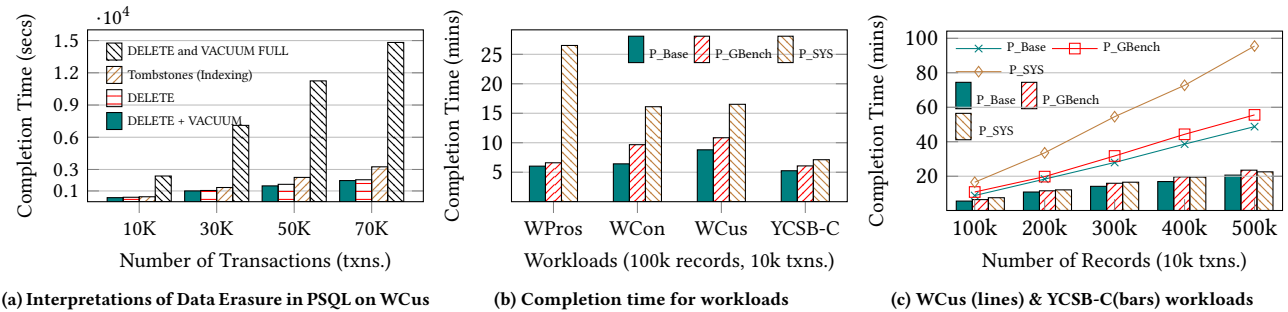
\section{Using Data-CASE}
\label{sec: Using Data-CASE}
We illustrate how we envision Data-CASE to be used using some experiments. We classify the uses based on the end-user. All our implementations were run on Oracle VM VirtualBox with a 6-cores (12 threads) AMD Ryzen 5 5600x 3.7GHz processor, 16GB RAM (DDR4-3200), and 50GB of disk space.
\subsection{Service Providers \& App. Developers}
\label{subsec: app dev}
Service providers and application developers often have their own system requirements which influence which database engine they use. Data-CASE offers a principled approach for them to identify desirable database properties and the corresponding data governance principles. This helps in choosing an appropriate data service provider which meets the desired requirements.

\noindent\textbf{\textsc{Case Study 1:}} Continuing our example, suppose MetaSpace, a service provider, wants to offer strong semantics of erasure to its customer to satisfy the requirements of $\mathcal{G}$17. They want to analyze which interpretations of erase can be supported by their database, PSQL, and their associated costs. To that end, they ground erase in Data-CASE (see Section \ref{sec: Grounding}) and identify system-actions \cite{DeletePostgreSQLDoc} offered by PSQL \cite{PostgreSQL} that can implement the groundings. We map the grounded erasure interpretations to system-actions supported by PSQL in Table \ref{tab:erasure_notions}. To assess how the implementation of each grounding impact system performance, they are benchmarked using the customer workload ($20\%$ deletes on data, rest are reads.) in GDPRbench \cite{shastri2019understanding}. The results are summarized in Figure \ref{fig:Eval_WkLds}(a). \textcolor{black}{VACUUM+DELETE surprisingly takes less time than only DELETE for the GDPRBench workload. VACUUM reclaims storage occupied deleted tuples that are not physically removed when only DELETE is used. The extra time taken by VACUUM in the deletion operations (20\%) is offset by the improved performance of the other operations (80\%) in the workload. Note that the expected performance is observed for a workload composed only of deletions.}
\subsection{Database providers}
\label{subsec: dbproviders}
Database providers like Oracle are often faced with the challenge of how to design databases that are compliant with data regulations or how to retrofit existing deployments to make them compliant. 
In Data-CASE, database providers can express concepts and actions supported in their existing deployments in terms of fundamental system properties or the effects the actions have on personal data. This fixes the interpretations of the concepts defined in Data-CASE. Now, a set of invariants is obtained that express the requirements for data governance. The system can then be retrofitted to meet those requirements it initially didn't.
When building new systems, Data-CASE lets the designer consider a wide range of interpretations, analyze their possible overheads, impacts on data- and control- paths, and expenses.

\noindent\textbf{\textsc{Case Study 2:}} 
Consider RelDB, which offers its services to various service providers. System designers at RelDB want to determine how to make their system, which runs on PSQL, GDPR compliant efficiently --- minimize costs and impact on system performance and at the same time offer meaningful interpretations useful for its clients. We show Data-CASE supports this.

Three interpretations of GDPR-compliance are implemented using associated groundings of concepts and by extending PSQL to support these groundings using system-actions. These are:\\ 
\noindent$1) $ P\_Base: 
The system implements role-based access control using roles, role attributes, and role memberships. It implements histories using native \texttt{csv} logging and setting up security policy to record query responses at row-level and the data is encrypted using AES-256 \cite{AES}. It implements deletes (see Table \ref{tab:erasure_notions} for grounding) to erase data using DELETE + VACUUM. The first interpretation of compliance is the least restrictive, and thus, is expected to have the least impact on system performance.\\
\noindent$2) $ P\_GBench: The system stores policies and other metadata in a table separate from the one containing personal data. Thus, all queries must perform joins to implement appropriate policies. Histories are implemented by logging all queries and responses (no csv logs). Data is encrypted using LUKS(SHA 256) \cite{LUKS, SHA}. Erasure is implemented using DELETE in PSQL.\\
\noindent$3) $ P\_SYS: The system implements fine-grained access control (FGAC) \cite{DBS-014}. Since PSQL does not support FGAC, it is retrofitted with a middleware that comprises Sieve \cite{Sieve} and associated metadata which implements FGAC  by exploiting a variety of its features such as UDFs, index usage hints, etc. to scale to a large number of policies. Data units and logs are encrypted using AES-128 \cite{AES} and erasure is implemented using DELETE + VACUUM FULL as well as deleting logs of the data units being deleted.

Evidently, the three systems have different interpretations of GDPR-compliance which changes the code and design. To validate and measure the expected varying amount of impact of the interpretations on the performance of these three systems, they are evaluated using the GDPRBench \cite{shastri2019understanding} and the industry-standard Yahoo Cloud Servicing Benchmark (YCSB) \cite{YCSB} (Workload-C). GDPRBench has three workloads namely  Controller[WCon] 
($25\%$ create, $25\%$ deletes, and $50\%$ updates to metadata), Processor[WPro] 
($80\%$ reads of data using keys, and $20\%$ reads of data using metadata), and Customer[WCus] 
($20\%$ each of reads, updates, and deletes of data, and reads and updates of metadata). We enriched the data records in GDPRBench with the Mall dataset from \cite{Sieve} comprising simulated data generated from personal devices in a shopping complex. Each record consists of a personal data-id and the recorded date and time generated using the SmartBench simulator \cite{gupta2020smartbench}. 
\color{black}

\textit{Metrics.} 
We analyzed 
{\em  completion time}, i.e., the total time taken to complete all the queries for each workload. 
To evaluate the ``Metadata explosion'' \cite{shastri13understanding} associated with each grounding/~implementation, we define \emph{space factor} as the ratio of the total size of the database to the total size of personal data in it. 

\textit{Summary of experiments and evaluation.} 
Figure \ref{fig:Eval_WkLds}(b) shows the overhead of implementations P\_Base, P\_GBench, and P\_SYS across the four workloads (each with 100k records and 10k transactions). In each case, the overhead of
P\_SYS is higher compared to P\_GBench which is higher than P\_Base. This is expected since the implementations use increasingly restrictive notions of compliance. 
The overheads in P\_GBench are small in comparison to those in P\_Base in WPro which consists of read queries. This small overhead is due to a slight increase in the information being logged. In contrast, since P\_SYS requires fine-grained policies to be checked, it incurs significant overhead. The difference between the completion times for P\_Base and P\_GBench in WCon
is larger compared to that in WCus and WPro due to a larger share of create, delete, and updates WCon. Such operations require more metadata access and logging. Likewise, the slowdown due to policies is more  profound in WPro in P\_SYS since it contains a larger number of read queries (100\% compared to others that are 50\% percent).
For WCon, P\_SYS still has a higher completion time compared to P\_Bench and P\_Base even though the workload is comprised of create, delete, and update and no reads which invoke expensive policy checks. However, all policies are logged at the time of all the operations to implement demonstrable accountability for logging requirements.

The YCSB Workload-C takes the least time to complete in each implementation since it does not require any associated metadata operations thereby highlighting the impact of changes required for compliance is small on non-GDPR operations.

Scalability was explored by increasing the volume of data but keeping the number of total transactions the same. Fig. \ref{fig:Eval_WkLds}(c) plots the completion time for the three systems running the WCus workload with an increasing number of data records and 10k transactions. As observed in earlier experiments, P\_Base takes the least time to complete the workload and P\_SYS takes the longest. As expected, P\_SYS is impacted the most by the increase in the size of data whereas the effect in P\_Base is the least. 
\begin{table}[h!]
    \centering
    \caption{Storage space overhead corresponding to Figure \ref{fig:Eval_WkLds}(b). 
    \footnotesize{The total size of P\_GBench and P\_SYS include indices.}}
    \small
    \begin{tabular}{ccccc}
    \hline
          \textbf{System}  & \textbf{Personal } & \textbf{Metadata} & \textbf{Total DB} & \textbf{Space} \\
           & \textbf{data size (MB)} & \textbf{size (MB)} &\textbf{size (MB)} & \textbf{factor} \\ \midrule
         P\_Base & $7$ & $14$ & $21$& $3\times$\\
         P\_GBench & $7$ &  $10$ & $26$& $3.7\times$\\ 
         P\_SYS &$7$ & $111$ & $120$& $17.1\times$\\ \hline
    \end{tabular}
    \label{tab:space_factor}
\end{table}
Groundings and their corresponding implementations impact database size. The size of PID remains the same (7 MB) but that of the metadata changes across interpretations (see Table \ref{tab:space_factor}). P\_GBench and P\_SYS use indices that occupy additional space. Recall that P\_SYS uses Sieve \cite{Sieve} which uses additional metadata. 

In summary, for entities like RelDB, our model provides a systematic approach to fix an interpretation of a data regulation and identify system-actions required to implement the interpretation and make design choices such as adding components, plugins, etc. to support chosen interpretations and study their characteristics. This paves the way to achieving demonstrable compliance.  

\subsection{Multinational organizations}
These often need to comply with conflicting and varying principles of data governance. GDPR itself allows EU member states to define their own data processing principles. Moreover, countries around the world have different data regulations. Entities are not ready to deal with the resulting complexities \cite{GDPRfailing}. 
Data-CASE supports varying interpretations of data regulations and makes the process of mapping data regulation requirements to precise system-actions transparent and unambiguous. Thus, it can help make decisions such as data geo-location, processors to use, and the consequences on services and features offered to its clients.
\subsection{Other Uses}
\label{subsec: other}
\noindent
\textbf{\textsc{Privacy Impact Assessment (PIA)}:} GDPR ($\mathcal{G}$35) imposes the burden of a PIA on controllers  prior to starting data processing. Such pre-deployment assessments of new, potentially high risk to the privacy and security of personal data, are often required by data regulations. Data-CASE supports impact assessments by providing system designers with system-actions (to implement specific groundings of the concepts) corresponding to each step in the data processing pipeline and their properties and interactions with each other. 
Once the risks have been identified and assessed, Data-CASE supports in implementing specific system-actions to mitigate those risks.
\\
\noindent
\textbf{\textsc{Regulatory Agencies}:} All data regulations establish regulatory agencies (e.g. see $\mathcal{G}$ 31) which certify that a data processing system is, indeed, compliant with that data regulation. For example, GDPR is enforced by Individual data protection authorities (DPAs) from the 27 EU member states. These agencies often have conflicting, non-transparent certifying processes and have repeatedly expressed frustrations with data regulations \cite{GDPRfrustration}. Data-CASE provides such agencies to identify groundings of concepts that are required, at minimum, to be in compliance with a data regulation. Conversely, agencies may require entities to use frameworks like Data-CASE to demonstrate the groundings of data processing adopted in a system and the system-actions that implement them.

\section{Related Work}
\label{sec: related work}
Conflicting priorities of data regulations and prevalent database systems \cite{shastri2019seven, shastri2021gdpr}, motivated preliminary studies in \cite{shastri2019understanding} which showed that GDPR-compliance severely impacts the performance of databases. Domain-specific work like \cite{GDPRNamedDataNet, kammueller2018formal} explores the consequence of GDPR in named data networking and Healthcare Systems, respectively. 
The consequences on policy and privacy management have been investigated in ~\cite{greengard2018weighing,wachter2017counterfactual,basin2018purpose}. 

Retrofitting databases to make them compliant has been explored in \cite{agarwal2021retrofitting, shastri2019understanding,davari2019access, Odlaw} and new, compliant-by-construction, systems have been proposed in \cite{schwarzkopf2019position, kraska2019schengendb, Mitra2009AnAF}.
Frameworks to implement
GDPR compliance have been explored in several prior works, especially in the 
context of data retention/erasure \cite{sarkar2022query, sarkar:hal-01824058, Lethe, scope2022purging} and policies. 
The work in \cite{mohan2019analyzing} explores 
privacy policies in large-scale cloud systems, \cite{ferreira2022rulekeeper} explores policy compliance in web frameworks,  \cite{tchana2022rgpdos} 
explores compliance
 in operating systems, \cite{Odlaw} builds a visual tool for managing data flow in systems,  while 
 \cite{AuditGaurd} explores auditing and retention policies in databases. A middleware layer to implement consent management \cite{daoudagh2021improve} and access control \cite{Sieve, tattletale} in databases have also been explored.

 Unlike such frameworks, our goal in this paper is to develop a simple 
 model for data and data processing that
 can be used to define GDPR/data regulation requirements 
 formally such that designers of systems
 such as the above can precisely define
 their interpretation of regulations and
 establish  compliance by illustrating that their software techniques indeed
 meet the formal requirements.  
In this sense, our vision is more 
related to prior approaches such as \cite{robaldo2020formalizing,robaldo2020dapreco,karami2022dpl}
that have explored formal logic-based GDPR specifications  that support verification of compliance through model checking, such as in \cite{TemporalModelCheck, compVeriIOT}. But these logic-based specifications at the level of broad data processing concepts remain vague from a system-compliance perspective. Unlike Data-CASE, these frameworks do not support the specification of how the concepts are interpreted or implemented in the system being verified. Such work complements Data-CASE with formal specifications potentially 
serving as \emph{invariants} in Data-CASE.

A rich line of work exists on modeling, specifying, implementing policies, and auditing  \cite{TemporalModelCheck, YoungGarg, amiri2022prever, automatingcomp, vacuuming, recordsretention, HarmonyPrivacy}. These can be a part of Data-CASE and the middlewares to audit such policies can support system-actions to maintain related invariants in Data-CASE. Privacy frameworks such as contextual integrity \cite{BarthNissen, ContextIntDataLeak} and origin privacy are related but orthogonal to our contributions. Our goal is not to create a new way of specifying what privacy should mean. Instead, given what privacy should mean, and data processing concepts grounded based on it, our framework’s goal is to provide ways to reason about whether a given system is compliant.
\color{black}
 Some compliance guidelines, specific to data regulations, are available from Governmental organizations, white papers, and blog posts \cite{team2020eu, AWSGDPRGuide, gdpr.eu, GoogleCloudGDPRComp,GoogleCloudPrivacyTerms} and offer some insights. 
\section{Challenges Ahead}
\label{sec: challenges}
GDPR violations (tracked by \cite{gdpr-tracker}) related to non-compliant systems continue to rise exponentially \cite{gdpr-tracker, CNILGoogle, MGDPRadopt}; the latest being that by Meta in May 2023 \cite{NYTFB23} which incurred the largest fine till date. 
Our paper makes a case that designing, deploying, and reasoning about the compliance of systems to data regulations like GDPR and others requires a formal framework to express data processing concepts. The paper proposes such a formal framework entitled Data CASE and shows how data governing principles of data regulations can be formalized in the form of grounding concepts defined into concrete system-actions and by defining invariants using these concepts. A full realization of our vision of a formal framework to support the development and deployment of compliant systems opens up several challenges:
\\
$\bullet $\textbf{ Completeness and correctness:} Demonstrating the correctness of a formal framework like Data-CASE and to what degree it can capture the system requirements of a data regulation remain unexplored. This is an important step towards compliance.
\\
$\bullet $\textbf{ Grounding concepts:} We focused on erasure and its possible interpretations in systems. Other concepts which capture the requirements of data regulations need to be defined. Once defined, 
special attention needs to be given to the interactions between and compatibility of different possible interpretations of the concepts.
\\
$\bullet $\textbf{ From a formal framework of compliance to a system to support compliance:} Data-CASE supports compliance-related decision-making. Automating this process will enable us to build a comprehensive tool that can be retrofitted on any non-compliant system to make it compliant with a given data regulation. Traditionally, retrofitting for compliance requires multiple tools and often complicates data flow \cite{davari2019access}.

In summary, we believe Data-CASE opens up a new direction of exciting research possibilities.
\begin{acks}
The first author was supported in part by the Irvine Initiative in
AI, Law, and Society and the HPI Research Center in Machine Learning and Data Science at UC Irvine. This work was supported by NSF Grants No. 2032525, 1545071, 1527536, 1952247, 2008993, 2133391, and 2245372. The authors thank the reviewers for their feedback.
\end{acks}
\bibliographystyle{ACM-Reference-Format}
\bibliography{reference}

%
% ...

\end{document}